\documentclass[twocolumn,amsmath,amssymb]{revtex4}

\usepackage{times}
\usepackage{graphicx}
\usepackage{bm}

\def\Vec#1{\boldsymbol {#1}}
\def\hsigma{\hat{\sigma}}
\def\hs{\hat{s}}
\def\TR{\mathrm {Tr}} 
\def\sgn{\mathrm{sgn}}

\begin{document}
\sloppy

\title{Naive Mean Field Approximation for the Error Correcting Code}
\author{Masami Takata}
\affiliation{Graduate School of Human Culture, Nara Women's University}

\author{Hayaru Shouno}
\affiliation{Faculty of Engineering, Yamaguchi University}
\email{shouno@ai.csse.yamaguchi-u.ac.jp}

\author{Kazuki Joe}
\affiliation{Graduate School of Human Culture, Nara Women's University}

\author{Masato Okada}
\affiliation{Brain Research Institute, RIKEN}

\date{\today}

\begin{abstract}
 Solving the error correcting code is an important 
 goal with regard to communication theory.
 %Recently, to reveal the characteristics of the error correcting code, 
 To reveal the error correcting code characteristics, 
 several researchers have applied a statistical-mechanical approach to this
 problem.
 In our research, we have treated the error correcting code as a Bayes
 inference framework. 
 Carrying out the inference in practice, 
 we have applied the NMF (naive mean field) approximation to 
 the MPM (maximizer of the posterior marginals) inference, 
 which is a kind of Bayes inference.
 %and, 
 In the field of artificial neural networks, 
 this approximation is used to reduce computational cost
 through the substitution of stochastic binary units with the deterministic
 continuous value units.
 However, %there exists  few report of the performance of this approximation quantitatively.
 few reports have quantitatively described the performance of this approximation.
 Therefore, we have analyzed the approximation performance 
 from a theoretical viewpoint, and
 have compared our results with the computer simulation.
\end{abstract}

\maketitle
\section{Introduction}
Within the framework of an error-correcting code, 
a sender encodes a message with redundant information added to
the transmitted sequence, and 
a receiver obtains the noise-corrupted sequence through 
a noisy transmitting channel.
Decoding process of the error-correcting code
is to restore 
the transmitted message 
from the received sequence that is corrupted through the transmitting process.
%
% モデルのこと
%

% 危険かも
Sourlas suggested that the error-correcting code can be dealt with
Bayesian inference, and proposed an encoding model
where all possible combinations of $r$-bits must be multiplied
as the redundant information
just like the Mattis model \cite{Mattis76}.
Sourlas also showed that the channel capacity of the model can reach
the theoretical limit, called the 
Shannon bound \cite{Shannon48} in the limit of $r\rightarrow\infty$.
Unfortunately, under the condition $r\rightarrow\infty$, the transmitting
speed becomes to $0$, so that the model with large $r$ is not practical.
%
% Kabashima らの成果を書くんだったらここかな。
%
Recently, however, Kabashima \& Saad pointed out that the Sourlas
encoding model with a small $r$, -- for example $r=2, 3$  --, is capable of 
good restoration ability with a practical transmitting speed \cite{Kabashima99}.
%
%
%
%
% References to Sourlas and Rujan
%

%
% 符号戦略のこと
%

In this paper, we also treat this problem according to 
a Bayesian inference framework.
This framework is based on estimating the restored message occurrence probability (posterior)
from both
the prior probability meaning the original message generation probability 
and the likelihood, which depends on the corruption process model.

% From: Nishimori \& Wong
%A general strategy common in error-correcting codes and image
%restoration is to use the Bayes formula on the posterior of an output
%sequence, given the input sequence.
%One then often accepts the sequence (image), which maximizes the
%posterior as the decoded/restored result.
%This method is called the maximum a posteriori probability estimate.
%
One strategy using Bayesian inference is to use
a message which maximizes the posterior probability.
This method is called maximum a posteriori (posterior) probability
(MAP) inference.
Given a corrupted sequence, the MAP inference accepts as the restored
result the message which maximizes the posterior probability.
From the statistical mechanical point of view, 
the logarithm of the posterior probability can be regarded as the energy, 
so we can consider the MAP inference as an energy minimization problem.
%
%Hence, the error-correcting code can be regarded as an optimization problem.
%
%
%
%The MAP estimation is a well-known method in the field of image restoration problem.
%Geman \& Geman demonstrated the MAP inference result applied to 
%the image restoration problem using the simulated annealing \cite{Geman84}, 
%which is a tool for searching the ground state \cite{Kirkpatrick83}.
%The image restoration problem is similar to the error correcting code 
%problem in the sense that
%the received bit sequence which corresponds to the image represented by
%a set of pixels is corrupted by noise, 
%and
%the receiver tries to retrieve the original bit sequence/image
%from the noisy one.
%The major difference between the error correcting code 
%and the image restoration is handling of the prior and 
%the redundant information embedded in the sequence.
%In the case of applying MAP inference to the error-correcting code, 
%we cannot assume any condition to the original message generating prior, 
%so that the uniform distribution is usually chosen for the prior.
%To compensate the prior information, 
%the redundant information about the message is embedded into the transmitting sequence.

%On the contrary, image restoration problem is usually given only the corrupted
%image not other additional redundant information.
%Therefore, in the image restoration, we usually assume 
%the alternative information for original image such as 
%the prior probability.

Another strategy is to use inference in which 
we consider the expectation value with respect to 
the maximized marginal posterior probability 
at each site's thermal equilibrium 
as the original message.
This method is called maximizer of the posterior marginals (MPM)
inference\cite{Rujan93}\cite{Sourlas94}\cite{Nishimori00}. 
From the statistical mechanical point of view, 
the MPM inference corresponds to minimization of the free energy.
In the MAP inference, 
the posterior probability is given for each candidate of sequences.
In contrast, in the MPM inference, 
the posterior marginal probability is given for each bit in the
sequence,
and the state which has the largest posterior probability is chosen as the restored state
for each bit.
Hence, to find a optimal bit sequence through the MPM inference, 
we should compare the posterior probability for each bit, and 
this requires that we calculate the thermal average for each sequence bit. 
%
% error correcting code and image restoration 
%
Recently, the MPM inference has been discussed with regard to error
correcting code \cite{Rujan93} \cite{Sourlas94} \cite{Nishimori00}.
Ruj\'an pointed out the effectiveness of 
carrying out a decoding procedure not at the ground state, 
but at a finite temperature\cite{Rujan93}.
Sourlas used the Bayes formula to re-derive the finite-temperature
decoding of Rujan's result under more general conditions \cite{Sourlas94}.
Finite-temperature decoding corresponds to the MPM inference, 
while decoding at the ground state corresponds to the MAP inference.
%
%
%
% Nishimori \& Wong らによる MAP < MPM 
%
%Nishimori \& Wong pointed out that the MPM inference involves the MAP inference:
%at the limit of the temperature $T\rightarrow 0$, 
%the MPM inference becomes equivalent to the MAP inference \cite{Nishimori00}.
Comparison of the restoration ability of the MPM inference to that of
the MAP inference
with regard to the error rate per bit has shown that,
the MPM inference is superior to the MAP inference \cite{Rujan93}\cite{Nishimori00}.
%
%evaluated by the thermal average in finite temperature $T>0$, 
%so this method is also called ``finite temperature restoration''.
%and the MPM inference is superior to the MAP inference 
%in the sense of restoration ability of each bit \cite{Nishimori00}.
%
%
%

\begin{figure*}[t]   %% try positioning first at the top, then bottom
\begin{center}
  \resizebox{0.8\textwidth}{!}
  {\includegraphics{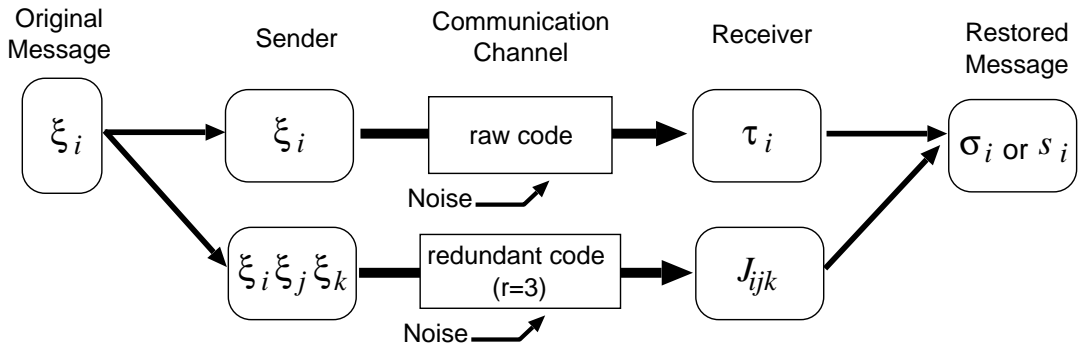}}
 \caption{Schematic diagram of a framework for the error correcting code.}
 \label{fig:method}
\end{center}
\end{figure*}

To consider the error correcting code from a statistical mechanical
perspective,
we denoted the message length, 
where the message is represented by a binary unit sequence, 
as $N$,
and assumed that each unit can take a binary state $\{-1,+1\}$.
The number of feasible message combinations in this case would  reach $2^N$,
making it hard to find a correct message among the numerous candidates.
Thus, to apply the MAP or MPM inference in a practical way, 
we usually have to adopt some kind of gradient descent algorithm, 
such as the Glauber dynamics,
despite a risk that the solution will be captured by local minima.
Moreover, applying the MPM inference may take more computational time 
than applying the MAP inference.
In the MAP inference, once the system reaches 
a macroscopic equilibrium state,
each pixel value can be properly determined with a probability of $1$.
On the other hand, when the system reaches
a macroscopic equilibrium state in the MPM inference, 
each unit value cannot be deterministicly decided
because the probabilities for each binary state have finite values 
in finite-temperature decoding.
Therefore, we need to calculate the thermal averages for each unit, 
and this requires many samplings.
In this paper, we discuss an approximation that replaces
the stochastic binary units with deterministic analog units which can
take $[-1,+1]$ continuously. 
In other words, we introduce deterministic dynamics into 
the MPM inference to avoid the need to sample for the thermal average.
In statistical mechanics, this approximation
is sometimes called the ``naive mean field (NMF)''
approximation \cite{Bray86}.
The purpose of the NMF approximation has usually been to
enable deterministic analog units to 
emulate the behavior of stochastic binary units.
%
%By Applying the NMFE, each restoring unit, 
%which is defined as stochastic binary unit,
%is replaced by a deterministic analog unit which can take $[-1,+1]$ continuously.
This approximation has been applied to several combinatorial
optimization problems,
such as the traveling salesman problem (TSP) \cite{Hopfield86}.

One important advantage of applying the NMF approximation is that the 
calculation cost is reduced since
we can avoid calculating the thermal average 
which requires many samplings in stochastic dynamics.
However, the NMF approximation has typically been applied as a mere
approximation  
in algorithm implementation,  
%so that we should evaluate the ability of NMFE.
%Unfortunately, 
so 
few researchers who have used the NMF approximation 
have investigated its quantitative ability - e.g., 
the approximation accuracy.
In this paper, we discuss the quantitative ability of the NMF approximation
in the MPM inference\cite{Shouno02}.

Nishimori \& Wong formulated the image restoration problem and the error
correcting code from the
statistical-mechanical perspective
by introducing mean field models for binary image restoration, 
and 
analyzed this model theoretically through the replica method.
We have applied the NMF approximation to the formulation, and analyzed the
model through the replica method in
the manner of Bray et al \cite{Bray86}.

In the field of neural networks, a network model using the NMF approximation is
sometimes called an analog neural network. 
Roughly speaking, 
Hopfield \& Tank's network is a result of applying the NMF approximation to 
the optimization network proposed by Kirkpatrick \cite{Kirkpatrick83}\cite{Hopfield86}.
Therefore, a Hopfield-Tank type network can be regarded as a kind of analog neural network.
The formulation of this kind of optimization network is very similar to 
that of the Sourlas encoding model with $r=2$.
In our study, we applied the NMF approximation to the Sourlas encoding model
with $r\geq 2$, 
which is called `r-body interaction' from a statistical-mechanical viewpoint, 
so that 
our formulation can be considered
a natural extension of a Hopfield-Tank type network with higher-order
dimension interactions.

In Sec. II, we formulate the error-correcting code using 
Bayes inference in the manner of Nishimori \& Wong's
formulation\cite{Nishimori00}. 
%Recently, from the statistical-mechanical point of view, 

In Sec. III, we compare the results between from our analysis to those
of a computer simulation.
Within the limits of the mean field approximation, 
our results agreed with the simulation results.

\section{Model and Analysis}
\subsection{Formulation of the Error Correcting Code}
In this section, to make this paper self-contained,
we explain the error correcting code in the manner of 
Nishimori \& Wong \cite{Nishimori00}.
Fig. \ref{fig:method} shows a schematic diagram of the error
correcting code.
On the sender side, the original signal is represented by $\Vec{\xi}$, 
and each element is a binary unit that takes binary states $\xi_i =
\{-1, +1\}$ for $i=1 \cdots N$.
The number of elements means the message length which can be denoted as
$N$.

Through the transmission channel, the signals being sent are degraded by
noise, 
so redundant information is needed to enable retrieval of 
the original message.
%Recently, 
The error-correcting code has been discussed
from a Bayesian inference point of view
\cite{Rujan93} \cite{Sourlas94} \cite{Nishimori00}.
In the manner of the Sourlas code \cite{Sourlas94}, 
the redundant message is produced from the $r$-units product:
\begin{equation}
 \xi_{i_1} \xi_{i_2} \cdots \xi_{i_r}, 
  \label{eq:redundant}
\end{equation}
where the indices satisfy $1 \leq i_1 < i_2 < \cdots < i_r \leq N$.
The sender transmits the redundant message $\xi_{i_1} \xi_{i_2} \cdots
\xi_{i_r}$ for all possible combinations of the indices.
Thus, the redundant message length becomes $_N C_r$.
In addition, we assumed that the original message $\{\xi_i\}$ has a 
uniform prior probability:
\begin{equation}
 P_s( \Vec{ \xi} ) = \frac{1}{2^N} \quad {\text{for any }}\Vec{\xi}.
  \label{eq:prior}
\end{equation}

On the receiver side, degraded signals are observed since
transmission channels ares noisy.
In this paper, received signals 
$\Vec{\tau}$ correspond to the original message $\Vec{\xi}$ 
(whose elements consists of $\xi_i \: (i=1 \cdots N)$)
and their elements are denoted as $\tau_i \: (i=1 \cdots N)$, 
while signals $\Vec{J}$ 
(whose elements consists of 
$J_{i_1 \cdots i_r} \: (1 \leq i_1 < \cdots < i_r \leq N)$)
correspond to the redundant message whose elements consists of 
$\xi_{i_1}\cdots \xi_{i_r}$.
The receiver should be able to estimate the original message $\Vec{\xi}$
from the received signals $\Vec{\tau}$, $\Vec{J}$.

To apply Bayesian inference to estimate the original message, 
we should consider the posterior probability based on observation: 
\begin{equation}
 P(\Vec{\xi} |\Vec{J}, \Vec{\tau}) = 
  \frac{
  P_{\mathrm{out}} (\Vec{J}, \Vec{\tau} | \Vec{\xi})  P_s( \Vec{\xi} )
  }
  {
  \TR_{\left\{ \Vec{\xi} \right\} } 
  P_{\mathrm{out}} (\Vec{J}, \Vec{\tau} | \Vec{\xi})  P_s( \Vec{\xi} )
  }
  \label{eq:posterior1}
\end{equation}
where $P_{\mathrm{out}} (\Vec{J}, \Vec{\tau} |\Vec{\xi})$ 
is a conditional probability of the observed signal which
can be regarded as the probability expression of the corrupting
process in the communication channel.

In this study, we assumed the communication channel is a Gaussian channel:
\small
\begin{align}
 P_{\mathrm{out}} (\Vec{J}, \Vec{\tau} | \Vec{\xi}) &=
 P(\Vec{J} | \Vec{\xi})  P(\Vec{\tau} | \Vec{\xi}) 
 \notag \\
 &= 
 \prod_{i_1 < \cdots < i_r} P(J_{i_1 \cdots i_r} | \Vec{\xi})
 \prod_{i=1} P(\tau_{i} | \Vec{\xi})
 \label{eq:noise1}
 \\
 P(J_{i_1 \cdots i_r} | \Vec{\xi}) &=
 \left( \frac{N^{r-1}}{J^2 \pi r!} \right)^{1/2}
 \notag\\
 &
 \times\exp \left(
 - \frac{N^{r-1}}{J^2 r!}  \left( 
 J_{i_1 \cdots i_r} - \frac{j_0 r! }{N^{r-1}} \xi_{i_1} \cdots \xi_{i_r}
 \right)^2
 \right)
 \\
 P(\tau_{i} | \Vec{\xi}) &=
 \frac{1}{\sqrt{2\pi \tau^2}}
 \exp \left(
 -\frac{1}{2\tau^2} 
 \left( 
 \tau_i - \tau_0 \xi_i
 \right)^2
 \right)
\end{align}
\normalsize
The first term in Eq.(\ref{eq:noise1}) corresponds to the noise of
the redundant message channel, 
and the second term corresponds to the noise of the original message channel.
The random variable
$J_{i_1 \cdots i_r}$  is i.i.d. according to a normal distribution 
whose mean is 
$\frac{j_0 r! }{N^{r-1}} \xi_{i_1} \cdots \xi_{i_r}$.
and whose variance
is $2 J^2 r!/N^{r-1}$. 
The random variable $\tau_i$ is i.i.d. according to a normal
distribution whose mean is $\tau_0 \xi_i$ and whose variance is $\tau^2$.
Thus, we can describe the transmission channel characteristics 
using the parameters $j_0$, $J$, $\tau_0$, and $\tau$.

%
% 事後確率を陽に書く
%
Substituting Eqs.(\ref{eq:prior}) and (\ref{eq:noise1}) into Eq.(\ref{eq:posterior1}),
we can denote the posterior probability as:
\small
\begin{align}
 & P(\Vec{ \xi} | \Vec{ J}, \Vec{ \tau}) =
 \notag\\
 &\quad
 \frac{1}{Z}
 \exp\left(
 \displaystyle 
 \frac{2j_0}{J^2} 
 \sum_{i_1 < \cdots < i_r} J_{i_1 \cdots i_r}
 \xi_{i_1} \cdots \xi_{i_r}
 +
 \frac{\tau_0}{\tau^2}\sum_{i=1} \tau_i \xi_i
 \right),
 \label{eq:posterior2}
\\
 & Z = 
 \TR_{\left\{\Vec{\sigma} \right\} }
 \exp\left(
 \displaystyle
 \frac{2j_0}{J^2} 
 \sum_{i_1 < \cdots < i_r} J_{i_1 \cdots i_r}
 \xi_{i_1} \cdots \xi_{i_r}
 +
 \frac{\tau_0}{\tau^2}\sum_{i=1} \tau_i \xi_i
 \right),
\end{align}
\normalsize
where $Z$ is a partition function.
%The receiver should estimate the original message from the received
%messages $\Vec{\tau}, \Vec{J}$, 
%and the restoring posterior probability is assumed by
%replacing the original message $\{\xi_i\}$ with the estimated message.
%In this paper, we treat deterministic analog units for decoding in the receiver restoration, 
%and compare the result with the previous work which adopted the
%stochastic binary units called Ising model\cite{Nishimori00}.
%Thus, we denote the estimated message by analog decoding as $\{\hs_i\}$,
%and by binary decoding as $\{\sigma_i\}$.

To distinguish the restored signal from the original one,
which is denoted $\Vec{\xi}$, 
we use the notation $\Vec{ \sigma}$ for the restored units.
Moreover, the noise channel characteristics,
denoted by $2j_0/J^2$, and $\tau_0/\tau^2$ in Eq..(\ref{eq:posterior2}), 
are not given on the receiver side, 
so the receiver should estimate these terms, 
which are called hyper-parameters.
We refer to describe these hyper-parameters as $\beta$, and $h$ respectively.
The posterior probability can thus be described as
\begin{align}
 & P(\Vec{ \sigma} | \Vec{ J}, \Vec{ \tau}) = 
 \notag\\
 &\quad
\frac{1}{Z}
 \exp\left(
 \displaystyle 
 \beta
 \sum_{i_1 < \cdots < i_r} J_{i_1 \cdots i_r}
 \sigma_{i_1} \cdots \sigma_{i_r}
 +
 h
 \sum_{i=1} \tau_i \sigma_i
 \right)
 \label{eq:posterior3} 
 \\
 & 
Z =  \TR_{\left\{\Vec{\sigma} \right\}}
 \exp\left(
 \displaystyle
 \beta
 \sum_{i_1 < \cdots < i_r} J_{i_1 \cdots i_r}
 \sigma_{i_1} \cdots \sigma_{i_r}
 +
 h
 \sum_{i=1} 
 \tau_i \sigma_i
 \right).
\end{align}

%
% そしてハミルトニアンへ
%
From  Eq. (\ref{eq:posterior3}), 
it is natural to introduce a Hamiltonian described as
\begin{equation}
 \beta H = -\beta \sum_{i_1 < \cdots < i_r} 
  J_{i_1 \cdots i_r} \sigma_{i_1} \cdots \sigma_{i_r}
  - h \sum_{i} \tau_i \sigma_i
  \label{eq:H1}.
\end{equation}
If we ignore the original message channel --, that is $h=0$ --, 
the Hamiltonian consists of only the r-body interaction term;
i.e., only the redundant message is effective for restoration, 
which is called the Sourlas code.
Sourlas introduced the Hamiltonian and discussed the macroscopic 
property of the error correcting code using statistical analysis
\cite{Sourlas94}.

%In this study, 
%we applied MPM inference
%, which is a kind of Bayesian inference, 
Rujan and Nishimori \& Wong have applied the MPM inference
to the error-correcting code as 
the restoration strategy \cite{Nishimori00}\cite{Rujan93}, 
that is adopting the $\sigma_i$ 
which maximizes the  marginal probability:
\begin{equation}
 P(\sigma_i | \Vec{J}, \Vec{\tau}) 
  = 
 \TR_{\left\{ \sigma_j \right\}_{j\neq i}}
 \frac{ 
 \exp( - \beta H ) 
  }
  {
  \TR_{\left\{ \sigma_j \right\}}
  \exp( - \beta H ) 
  }
 \label{eq:marginal1}
\end{equation}
In this case, this is equivalent to adopting $\sigma_i$ 
as
\begin{align}
 \sigma_i &= 
 \sgn\left( 
 P(\sigma_i = +1 | \Vec{J}, \Vec{\tau} ) -
 P(\sigma_i = -1 | \Vec{J}, \Vec{\tau} )
 \right)
 \notag \\
 &= 
 \sgn\left(
 \TR_{\sigma_i} \sigma_i P(\sigma_i | \Vec{J}, \Vec{\tau} )
 \right)
 \notag\\
 &=
 \sgn\left(
 \frac{
 \TR_{\left\{ \sigma_j \right\}}
 \sigma_i \exp( - \beta H ) 
 }
 {
 \TR_{\left\{ \sigma_j \right\}}
 \exp( - \beta H ) 
 }
 \right)
 \\
 &= \sgn \langle \sigma_i \rangle_{\beta}
 \label{eq:finite}
\end{align}
The term $\langle \sigma_i \rangle_{\beta}$ in Eq. (\ref{eq:finite})
represents the thermal average of $\sigma_i$  
in the Hamiltonian $H (in (\ref{eq:H1}))$ with the finite temperature $\beta$.
Therefore the MPM inference is called finite-temperature decoding.

%
% ここでMPMが有限温度修復に連関することを言う
% そしてオーバーラップの意味では MPM > MAP であること
%
Nishimori and Wong compared the MPM and MAP inferences
which used the $\Vec{\sigma}$ that minimized the 
Hamiltonian $H$ in Eq.(\ref{eq:H1}) \cite{Nishimori00}.
They pointed out that the MAP inference is equivalent to 
the MPM inference within the limit of the temperature 
$T (= \beta^{-1}) \rightarrow 0$.
They also suggested that 
the MPM inference is superior to the MAP inference
with regard to the overlap:
\begin{align}
 M_o &= 
 \left[ \frac{1}{N} 
 \sum \: \xi_i \: \sgn \langle \sigma_i \rangle_{\beta} 
 \right]
 \label{eq:ovlp1}
 \\
 & =
 \TR_{\{ \Vec{\xi} \}, \{ \Vec{J} \}, \{ \Vec{\tau} \}}
 P_s(\Vec{\xi}) 
 P_{{\mathrm{out}}} (\Vec{J}, \Vec{\tau}| \Vec{ \xi})
 \: \xi_i
 \: \sgn \langle \sigma_i \rangle_{\beta}.
\end{align}
In Eq.(\ref{eq:ovlp1}), the bracket $[\cdot]$ 
, which is called as the configuration average,
 denotes the averages over the sets $\{\Vec{\xi}\}$, $\{ \Vec{J} \}$, and $\{\Vec{\tau}\}$.
Restoration ability measured in terms of overlap is maximized
when the parameters $\beta$ and $h$ equal, respectively,  $2j_0/J^2$ and
$\tau_0/\tau^2$ \cite{Nishimori00}.

%
% でも MPM 時間かかるよ?
% アナログデコーディング
%
In the MPM inference, each restored unit $\sigma_i$, which is a stochastic binary unit, 
is subject to thermal fluctuation since the decoding is carried out at a finite temperature;
this means each $\sigma_i = \pm 1$ state has finite probability value.
Therefore, we should calculate the thermal averages for all of the units 
when decoding is carried out.
In contrast, 
decoding using the MAP inference is done at the temperature $\beta^{-1}=0$, 
so there is no thermal fluctuation does not occurred.
When the MPM inference is applied to information processing, 
the calculation cost of the thermal averaging must be evaluated,
and the averaging process requires a lot of calculation cost\cite{Shouno02}.
To avoid this high cost, we have introduced the naive mean field (NMF) approximation.
%
%NMF approximation means substituting stochastic binary unit $\sigma_i$
%into 
%the deterministic analog unit $s_i$, and
%assuming the thermal average $\langle \sigma_i \rangle$ as the output of
%$s_i$.
%
We previously reported that
an image-restoration model using the MPM inference with stochastic binary
units requires more than $50$ times as much computational time as a
model using the NMF approximation \cite{Shouno02}.

%
% stochastic Ising <--> binary
% deterministic Analog  <--> NMF
% をどこで書くべきか ?
%

\subsection{Naive Mean Field Approximation}
In this study, in an attempt to find the ground state in a practical way, 
we introduced the NMF approximation \cite{Hopfield86}\cite{Shouno02}.
When the NMF approximation is used, 
each stochastic binary unit $\sigma_i$ 
is replaced by an analog unit $s_i$ that can take a continuous value $[-1,+1]$, 
and the output of each analog unit is regarded as $\langle \sigma_i
\rangle$;
that is,  the thermal average of the corresponding binary unit output $\sigma_i$
can take binary states $\sigma_i=\pm 1$ stochastically.

To replace the stochastic binary units with deterministic analog units on
the receiver side, 
we introduced a Hamiltonian for substituting Eq.(\ref{eq:H1}).
%
% アナログハミルトニアンを導入
%
\begin{equation}
 \beta {\cal H} = - M
  \left( 
   \beta
   \sum_{i_1 < \cdots < i_r } J_{i_1 \cdots i_r}		    
   \hs_{i_1} \cdots \hs_{i_r}
   +
   h
   \sum_{i} \tau_{i} \hs_{i}
  \right) ,
  \label{eq:H2}
\end{equation}
\begin{figure}[tb]   %% try positioning first at the top, then bottom
\begin{center}
 \resizebox{0.45\textwidth}{!}
 {\includegraphics{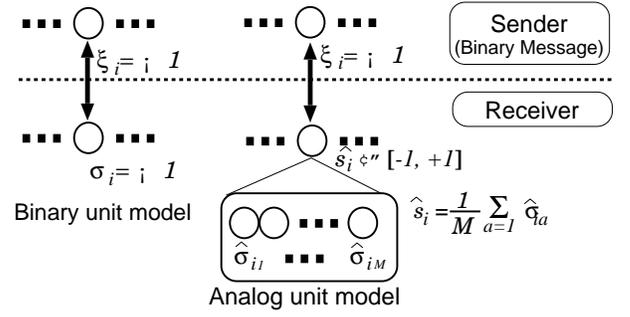}}
  \caption{Analog model for analysis:
 In the binary model, encoding unit $\xi_i$ corresponds to decoding unit
 $\sigma_i$.  
 In contrast, the analog model assumes a decoding unit has $M$ units, and
 the average of these $M$ units is regarded as the analog unit output.
 }
 \label{fig:fig2}
\end{center}
\end{figure}
where $\hs_i$ denotes an output of an analog unit which takes a 
continuous value $[-1,+1]$ in the equilibrium state.
$M$ is a scaling factor described below.
To analyze the model described by the Hamiltonian Eq.(\ref{eq:H2}), 
we followed the manner of Bray {\it et al.} \cite{Bray86}.
We assumed each $i$th site consists of $M$ binary units,
and the analog unit output $\hs_i$ can be calculated using the average of $M$
binary units $\hsigma_{ia}$ (see Fig.\ref{fig:fig2}).
\begin{equation}
 \hs_i = \frac{1}{M} \sum_{a=1}^{M} \hsigma_{ia}
\end{equation}
For the limit $M\rightarrow \infty$, each output $\hs_i$ can
take a continuous value $[-1,+1]$. When $M$ is a finite value, 
each $\hs_i$ is called a `binominal spins' which can take
$-1, -1+\frac{2}{M}, \cdots, 1-\frac{2}{M}, 1$ with a binominal distribution.
%
% グラウバダイナミクスの導入
%
We can thus introduce a `spin weight function' as
\begin{align}
 W(\hs_i) &= \:\: \TR \:\: \delta \left( M\hs_i - \sum_{a=1}^{M} \hsigma_{ia}
 \right) 
 \notag\\
 &= \frac{1}{2\pi {\mathrm j}} 
 \int_{-{\mathrm j} \infty}^{+{\mathrm j}\infty} du_i 
 \exp( M (-u_i \hs_i + \ln 2 \cosh(u_i) )).
\end{align}
The partition function $\cal Z$ can be described as
\begin{align}
 & {\cal Z} = \prod_{i=1}^{N} \int_{-1}^{+1} d\hs_i W(\hs_i) \exp(- \beta {\cal H} ) 
 \notag\\
 &=
 M  \int_{-{\mathrm j} \infty}^{+{\mathrm j}\infty} 
 \prod_{i=1}^{N} \left(\frac{du_i}{2\pi {\mathrm j}}\right)
 \int_{-1}^{+1} \prod_{i=1} d\hs_i
 \notag\\
 &\quad\quad
 \exp\Biggl( M \biggl(
 \beta
 \sum_{i_1 < \cdots < i_r } J_{i_1 \cdots i_r}		    
 \hs_{i_1} \cdots \hs_{i_r}
 \notag\\
 & \quad
 +
 h
 \sum_{i} \tau_{i} \hs_{i}
 -u_i \hs_i + \ln 2 \cosh(u_i) \biggr)\Biggr) 
 \label{eq:analogZ}
\end{align}
At the limit $M\rightarrow \infty$, the integrals over $\{u_i\}$ and
$\{\hs_i\}$ in Eq. (\ref{eq:analogZ}) can be evaluated by the
saddle-point method.
To derive the saddle-point equations, 
we differentiated the exponent of Eq.(\ref{eq:analogZ}) by $\hs_{i}$
and $u_i$, and obtained
\begin{align}
 0 &=  \frac{\beta}{(r-1)!} \sum_{i_2=1}^{N} \cdots \sum_{i_r=1}^{N}
 J_{i\: i_2 \cdots i_r} 
 \hs_{i_2} \cdots \hs_{i_r}
 + h \tau_{i}
 - u_{i},
 \label{eq:saddle1}
 \\
 0 &= -\hs_i + \tanh u_i.
 \label{eq:saddle2}
\end{align}
To derive Eq.(\ref{eq:saddle1}),
we assumed $J_{i_1 \cdots i_r}$ was symmetric:
\[
 J_{i_1 \cdots i_r} =  J_{i'_1 \cdots i'_r},
\]
where the indices group 
$(i'_1 \cdots i'_r)$ is any permutation group of  $(i_1 \cdots i_r)$ .
Moreover, we assumed 
self-coupled term in $J_{i_1 \cdots i_r }$ equals zero:
\[
 J_{i_1 \cdots l \cdots l \cdots i_r} = 0.
\]
Eliminating $u_{i}$ from Eqs.(\ref{eq:saddle1}) and
(\ref{eq:saddle2}), 
we obtain 
\begin{align}
 \hs_{i} &= 
 \tanh \left(  
  \frac{\beta}{(r-1)!}
 \sum_{i_2=1}^{N} \cdots \sum_{i_r =1}^{N}
 J_{i\: i_2 \cdots i_r} \hs_{i_2} \cdots \hs_{i_r}
  + h \tau_{i} 
 \right).
 \label{eq:equil1}
\end{align}
For example, for $r=3$, Eq.(\ref{eq:equil1}) can be denoted as
\begin{equation}
 \hs_i =
  \tanh \left( \frac{\beta}{2} \sum_{j=1}^{N} \sum_{k=1}^{N} J_{ijk} \: \hs_j \hs_k
	 + h \tau_i \right),
\end{equation}
where we assumed
$J_{ijk} = J_{kij} = J_{jki} = J_{kji} = J_{ikj} = J_{jik}$, and
$J_{ijj} = J_{jij} = J_{jji} = 0$.
%
%

%In the manner of Hopfield \& Tank \cite{Hopfield86}, 
We can derive naturally a discrete synchronous updating rule:
\begin{equation}
 s_{i}^{t+1} = 
 \tanh \left(  
  \frac{\beta}{(r-1)!}
 \sum_{i_2, \cdots, i_r}  J_{i \: i_2 \cdots i_r} \hs_{i_2}^{t} \cdots \hs_{i_r}^{t}
  + h \tau_{i} 
 \right).
 \label{eq:dynamics}
\end{equation}
where $s^{t}_i$ denotes the analog unit output at time $t$.
Eq.(\ref{eq:dynamics}) does not include any stochastic calculation,
so we refer to Eq.(\ref{eq:dynamics}) as the deterministic dynamics.
When the model described by Eq. (\ref{eq:dynamics}) 
reached to the equilibrium state $s^{\infty}_i$, 
all units should satisfy Eq. (\ref{eq:equil1}).
Therefore, to investigate the equilibrium state of dynamics
(Eq. (\ref{eq:dynamics})), 
we should use the analog Hamiltonian described by Eq.(\ref{eq:H2}).
From the idea of the NMF approximation,
each analog unit state expressed by
$s^{\infty}_i$ in the equilibrium state will   
correspond to $\langle \sigma_i \rangle$, i.e.
$s^{\infty}_i$ can be regarded as the thermal average of $\sigma_i$.
From Eq. (\ref{eq:dynamics}), this model follows the deterministic dynamics, 
so there is no need to calculate the thermal average of a stochastic unit; 
the expected calculation cost is thus lower than that for the stochastic binary units model.

\subsection{Equilibrium State Analysis}
%
% レプリカ法
%
To investigate the macroscopic property, 
we analyzed the NMF approximated model that includes the Hamiltonian described by
Eq. (\ref{eq:H2}) through the ``replica method'',
which is a standard statistical-mechanical analysis tool.
The MPM inference corresponds to the minimization of the free energy
denoted as $T [\ln {\cal Z}]$.
Here, ${\cal Z}$ is the partition function:
\begin{equation}
 {\cal Z} = \TR_{\{\sigma_{ia}\}} \exp( -\beta {\cal H}),
\end{equation}
where ${\cal H}$ is described as Eq.({\ref{eq:H2}}).
However, it is impossible to evaluate $[\ln {\cal Z}]$ in practical, 
we use replica trick $[\ln {\cal Z}] = \lim_{n\rightarrow 0} ([{\cal Z}^n]-1)/n$.
The $n$ replicated partition function $[{\cal Z}^n]$ can be expressed as:
\small
\begin{align}
 &\left[ {\cal Z}^n \right] 
 =
 \TR 
 \left(
 \int \prod_{i_1 < \cdots < i_r} \sqrt{\frac{N^{r-1}}{J^2 \pi r!}}
 dJ_{i_1 \cdots i_r} 
 \right)
 \left(
 \int \prod_{i=1} \frac{1}{\sqrt{2\pi\tau^2}} d\tau_i
 \right)
 \notag\\
 &\quad\quad
 \times
 P_s(\{\xi_i\}) 
 P_{{\mathrm{out}}} (\{J_{i_1 \cdots i_r}\}, \{ \tau_i \}| \{ \xi_i\})
 {\cal Z}^n, \\
%
% &=
% \TR
% \left(
% \int \prod_{i_1 < \cdots < i_r} \sqrt{\frac{N^{r-1}}{J^2 \pi r!}}
% dJ_{i_1 \cdots i_r} 
% \right)
% \left(
% \int \prod_{i=1} \frac{1}{\sqrt{2\pi\tau^2}} d\tau_i
% \right)
% \notag\\
% & \quad
% \times
% \frac{1}{2^N}
% \exp \left(
% - \frac{N^{r-1}}{J^2 r!}  
% \sum_{i_1 < \cdots < i_r}
% \left( 
% J_{i_1 \cdots i_r} - \frac{J_0 r!}{N^{r-1}} \xi_{i_1} \cdots \xi_{i_r}
% \right)^2
% -\frac{1}{2\tau^2} 
% \sum_{i=1}^{N}
% \left( \tau_i - \tau_0 \xi_i \right)^2
% \right)	
% \notag\\
% & \quad
% \times
 &
 {\cal Z}^n = 
 \prod_{\alpha=1}^{n}
 \exp\biggl(
 \frac{\beta}{M^{r-1}} \sum_{i_1 < \cdots < i_r } 
 J_{i_1 \cdots i_r} 
 \sum_{k_1,\cdots,k_r} 
 \hsigma_{i_1 k_1}^\alpha \cdots \hsigma_{i_r k_r}^\alpha 
\notag
\\
 & \quad\quad\quad
 +
 h \sum_{i} \tau_i \sum_{k} \hsigma_{ik}^{\alpha} 
 \biggr),
\end{align}
\normalsize
where operator $\TR$ in the last equation represents the sum over all states about
$\{\hsigma_{ia}^{\alpha}\}$ and $\{ \xi_i \}$.
We analyzed this replicated partition function through the standard replica
method.
The replica symmetry solution can be described as
\begin{align}
 m &= \int Dz \: G(z), \\
 q &= \int Dz \: G(z)^2,  \\
 \chi &=
 \frac{1}
 {\displaystyle \left(\frac{\beta^2 J^2 r q^{r-1}}{2} + h^2 \tau^2 \right)^{1/2}}
 \int Dz \: z G(z), 
\end{align}
where
\begin{align}
 & G(z) = \tanh \Biggl(
	       \biggl(
		\frac{\beta^2 J^2 r q^{r-1}}{2}
		+ h^2 \tau^2
	       \biggr)^{1/2} z
 \notag \\
 &\quad
	       +\frac{\beta^2 J^2}{2} r (r-1) \chi q^{r-2} G(z)
	       +\beta j_0 r m^{r-1} + h\tau_0 
	      \Biggr)
 \label{eq:G}
\end{align}
Using these solutions, we could obtain the overlap $M_o$:
\begin{align}
 M_o &= 
 \left[ 
 \frac{1}{N} \sum_{i=1}^{N} \xi_i \: \sgn( \hs_i ) 
 \right]
 \\
 &=
 \int Dz \: \sgn\left( G(z) \right)
\end{align}
If we Assume $\chi=0$, this analysis agrees with the result obtained 
using stochastic binary units reported by Nishimori\& Wong \cite{Nishimori00}.
Thus, in our analysis, the difference in the result that arises from 
the reaction term $\frac{\beta^2 J^2}{2} r (r-1) \chi q^{r-2} G(z)$ 
in Eq.(\ref{eq:G}).

\begin{figure*}[t]   %% try positioning first at the top, then bottom
\begin{center}
 \resizebox{0.8\textwidth}{!}
 {\includegraphics{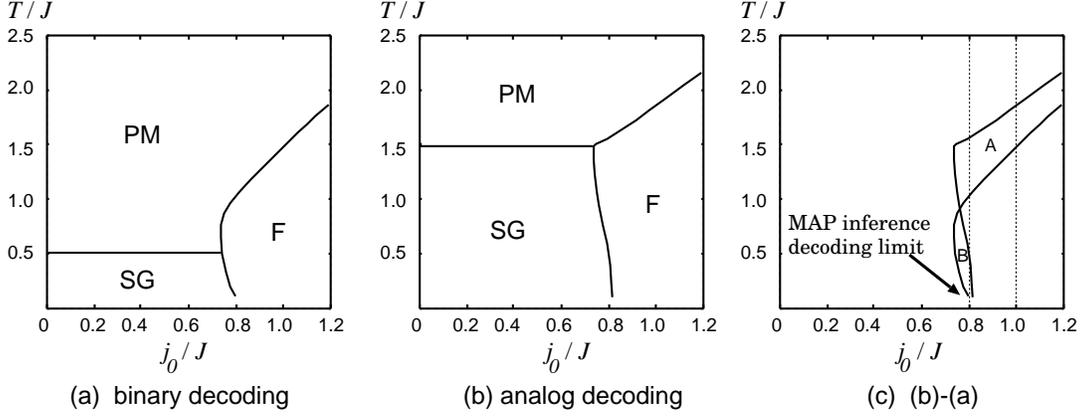}}
 \caption{
 Phase diagram of binary and analog models:
 The horizontal axis shows S/N ratio $j_0/J$, and the vertical axis
 shows the decoding temperature $T/J$.
 (a) shows the analysis result for the binary model, and (b) shows the
 result for the analog model.
 Region `F' shows the retrievable area.
 (c) shows the difference between (a) and (b).
 Region `A' is where  the analog model is superior to the binary
 model and the region `B' indicates the opposite.
 }
 \label{fig:phase}
\end{center}
\end{figure*}

\section{Result}
In this section, we compare the theoretical and simulation results 
for the conventional stochastic binary model and the NMF approximated
(deterministic analog)  model.
We refer to the decoding using the MPM inference 
with conventional stochastic binary units as the `binary model' and 
to the NMF approximated model as the `analog model'.
In subsection \ref{sec:result1}, 
we show the analytical results for each model.
In subsection \ref{sec:result2}, 
we compare the results from these theoretical analyses with those from computer simulations.
To compare these results, 
we configured an environment described by several parameters having the same value;
that is, $r = 3$, $h = 0$, and $J = 1$.
The condition $r = 3$ means a redundant message was generated by 3-body
$\xi_i \xi_j \xi_k$ where $1\leq i < j < k \leq N$.
We use $h = 0$, which means the decoding was carried out using only
information $\Vec{J}$ corresponding to the redundant message.
Thus, $\Vec{\tau}$ had no effect in our decoding models.
%And $J=1$ means the variance of the transmission channel is constant.
Under this condition, the original message was generated by uniform prior
probability Eq.(\ref{eq:prior}).
The redundant messages were generated by Eq.(\ref{eq:redundant}), 
and the message corruption process is described by Eq.(\ref{eq:noise1}).

In the simulation, to decode the original message $\Vec{\xi}$ from the corrupted signals $\Vec{J}$ 
using the stochastic binary model, 
we used a kind of gradient descent algorithm --, that is Glauber dynamics, --
to find the minimum state of the Hamiltonian $H$ described in Eq.(\ref{eq:H1}).
For decoding using the analog model, 
we used the update rule describe in Eq.(\ref{eq:dynamics}) to find the
minimum of the free energy of 
the Hamiltonian ${\cal H}$ described in Eq.(\ref{eq:H2}).

\subsection{The ability of NMF Approximation}
\label{sec:result1}
%
% ここでは NMFE で近似した奴と そうじゃない奴の差を
%
Fig.\ref{fig:phase}(a)  is a  phase diagram for the binary model with $r=3$.
In the figure, we only show the `replica symmetry' solution; a more
detailed solution has been given by Nishimori \& Wong \cite{Nishimori00}.
The $x$-axis shows $j_0/J$, which corresponds to 
the S/N ratio in the communication channel, and the $y$ axis shows
$T/J$, which corresponds to the decoding temperature.
Region `F', 
which represents the ferromagnetic state ($m>0$ and $q>0$),
in the figure shows the retrievable region.

Fig.\ref{fig:phase}(b) is a phase diagram for the
analog model using the same parameters as for the binary model.
Comparing figs.\ref{fig:phase}(b) and \ref{fig:phase}(a), 
we see that the decoding ability of the analog model looks better than 
that of the binary model in the high temperature region.
%To emphasize the difference between 
%the analog deterministic system
%and 
%the binary system, 
%we show the subtract the retrievable area of the binary stochastic system shown in 
%Fig.\ref{fig:phase_bin} 
%from 
%that of the deterministic analog system shown in Fig.\ref{fig:phase_ana}.

%In Fig.\ref{fig:phase_sub} which shows the subtract result,
%the area marked as `+' where deterministic analog decoding is superior.
%On the contrary, the `-' marked ared shows the binary stochastic
%decoding is superior,
%that is around $j_0/J =0.8$.
In Fig.\ref{fig:phase}(c), region ``A'' shows where the analog model
is superior to the binary model,
while region `B' shows where the binary model provides better decoding ability 
than the analog model.
The binary model is better when the S/N ratio is low, 
while the analog model is more robust with respect to the decoding temperature $T$.

\subsection{Comparison with Computer Simulation}
\label{sec:result2}
\begin{figure*}[t]
 \begin{tabular}{cc}
  \resizebox{0.45\textwidth}{!}
  {\includegraphics{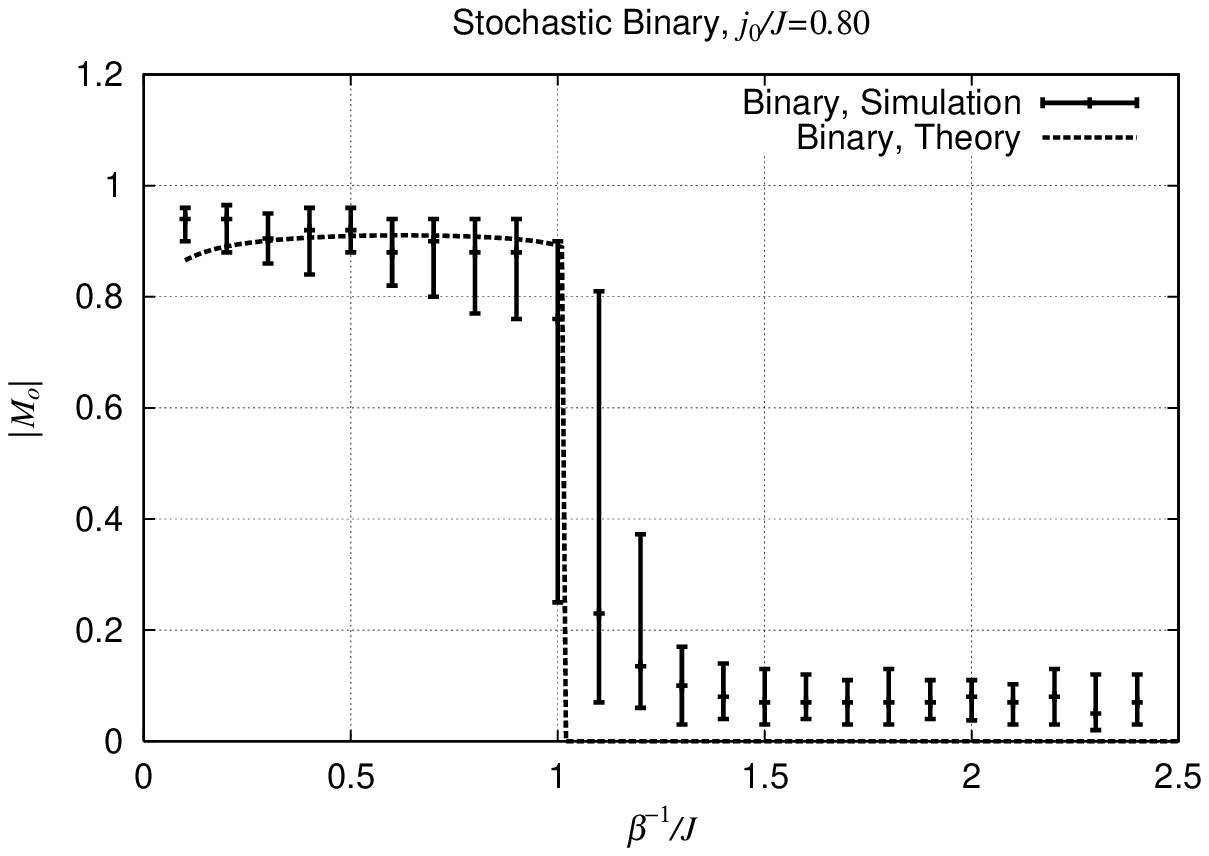}}
  &
  \resizebox{0.45\textwidth}{!}
  {\includegraphics{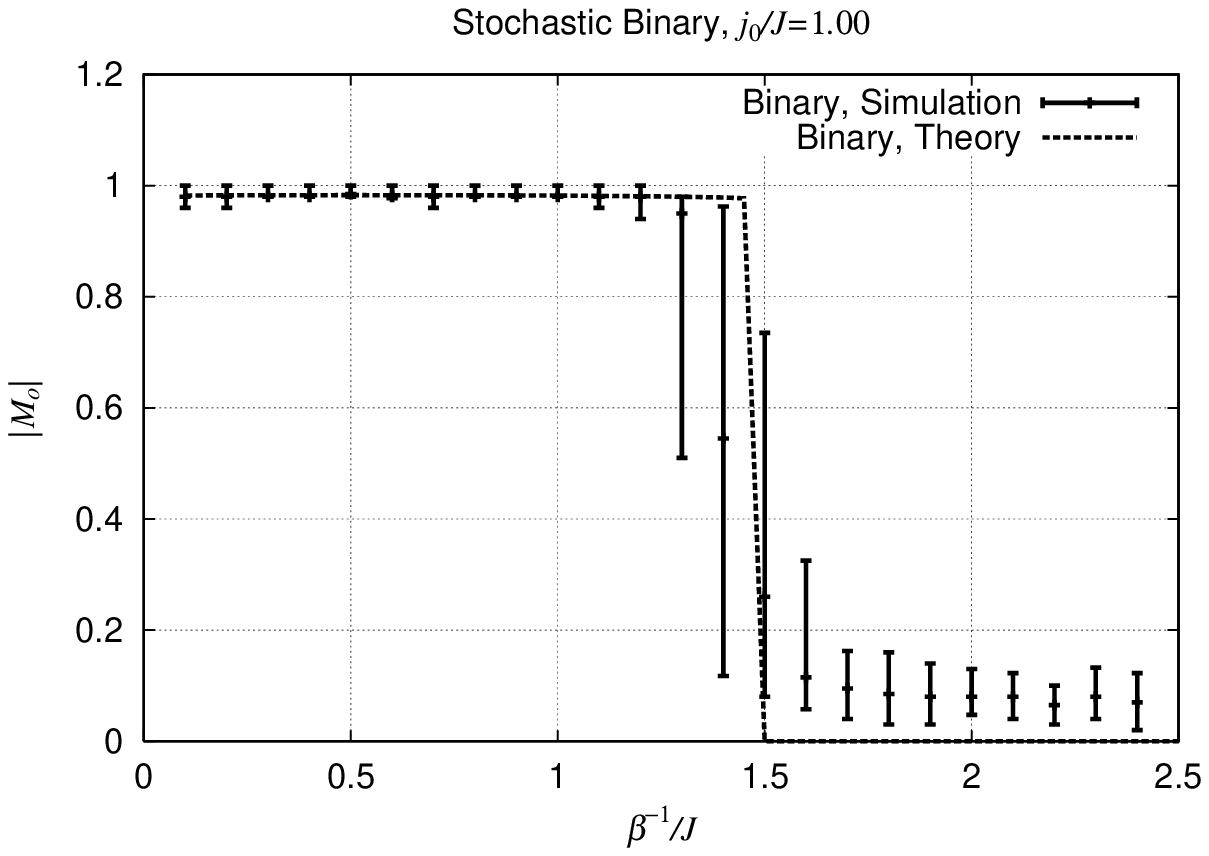}}
  \\
  (a) Binary, $j_0/J=0.80$, $N=100$  &   (c) Binary, $j_0/J=1.0$, $N=100$\\
  \resizebox{0.45\textwidth}{!}
  {\includegraphics{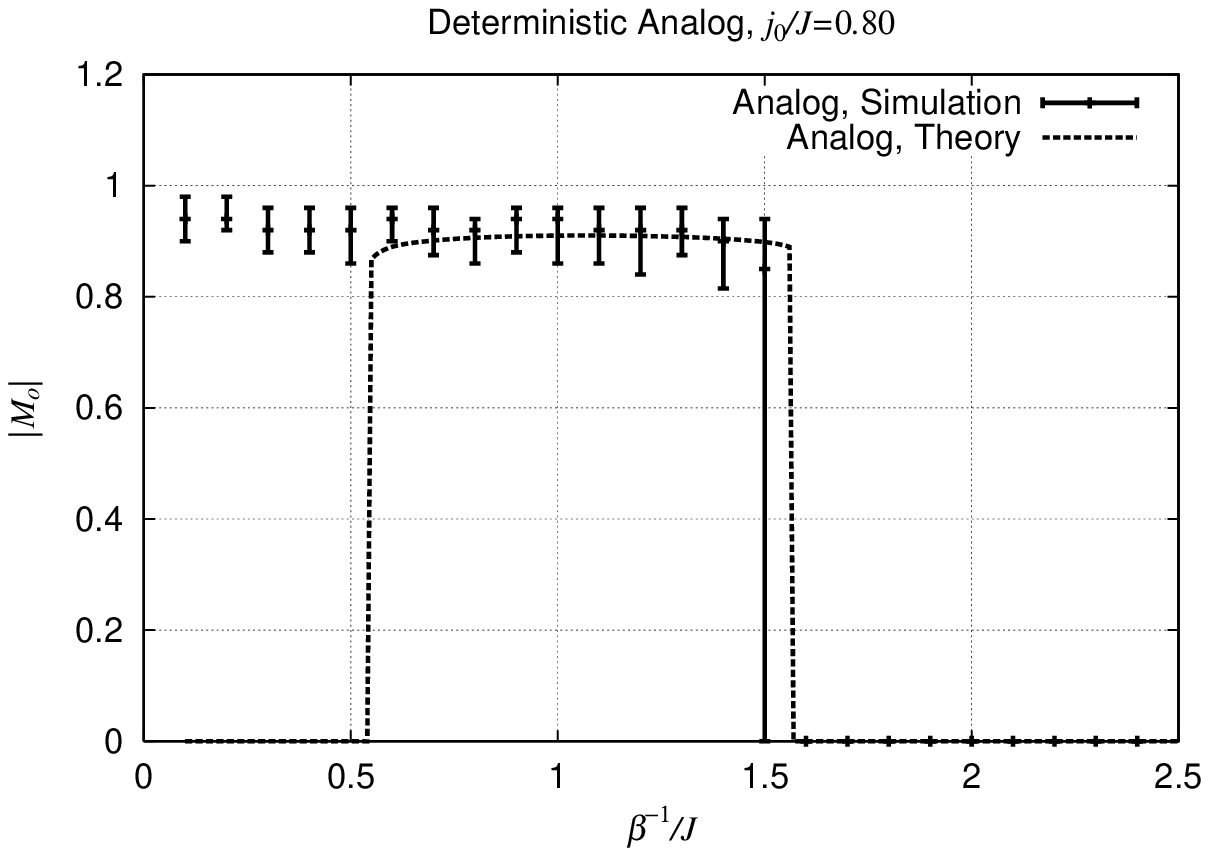}}
  &
  \resizebox{0.45\textwidth}{!}
  {\includegraphics{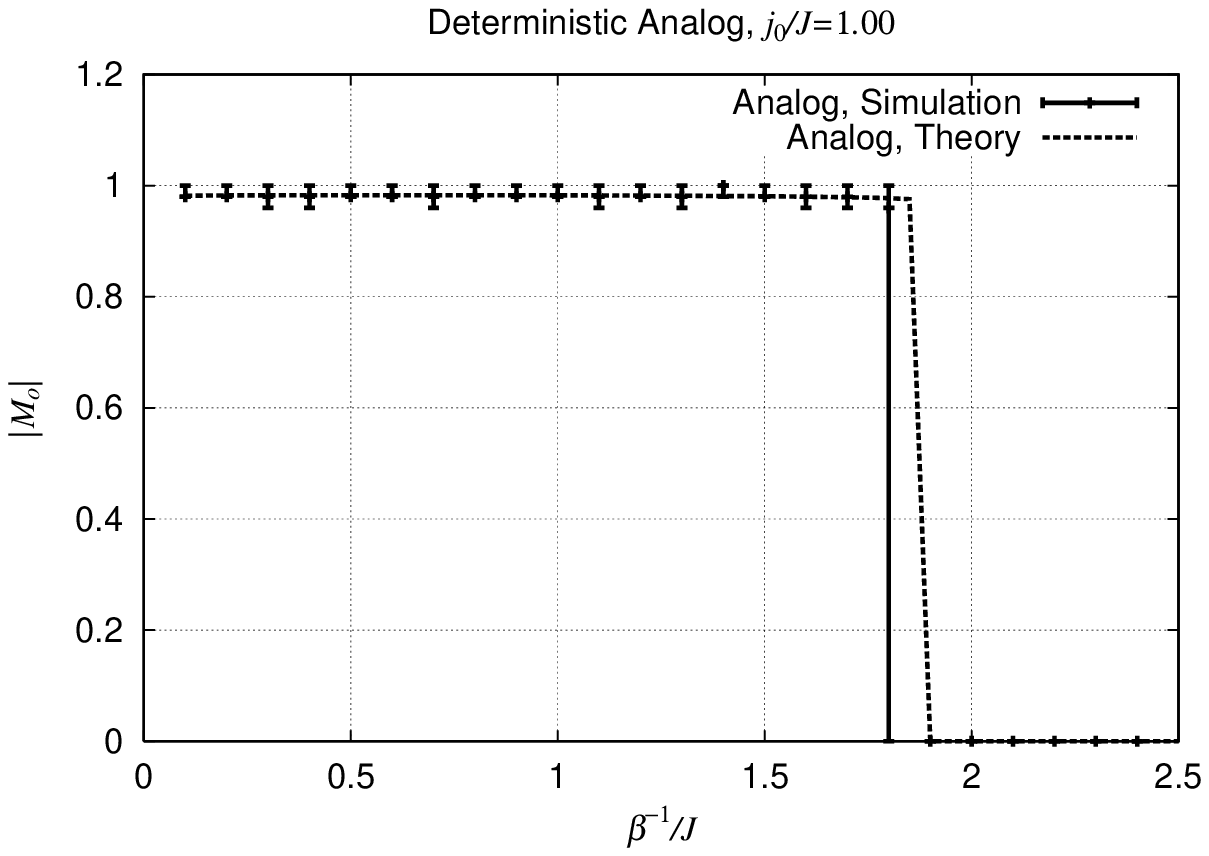}}
\\
  (b) Analog, $j_0/J=0.80$, $N=100$ &  (d) Analog, $j_0/J=0.80$, $N=100$
 \end{tabular}
 \caption{Comparison with Computer simulation. The error bars in each
 figure represent the computer simulation results. The horizontal axis shows
 the decoding temperature and the vertical axis shows  the absolute
 value of the overlap  $M_o$.
 Each error bar range represents the first and third quartile deviations, 
 and the middle point indicates the median.
 The dashed line in each figure shows the replica analysis result.
 }
 \label{fig:SimResult}
\end{figure*}

The theoretical results described in sec.\ref{sec:result1},
were realized in an equilibrium state, 
so we confirmed the validity of our theory through the computer simulations.
In the simulations, we assumed a message length as $N=100$.

%To confirm the analysis validity, we compared the result obtained by the
%theory with the computer simulation result.

In the binary model simulation,
we used asynchronous Glauber dynamics.
First, 
we randomly selected one site $l$, and 
calculated the transition probability as determined by the heat-bath method:
\begin{align}
 W( \Vec{\sigma}^t \rightarrow {\cal F}_l (\Vec{\sigma}^t)) =
 \frac{1}
 {1+\exp(\beta (H(\Vec{\sigma}^t) - H({\cal F}_l (\Vec{\sigma}^t) ) ) )},
\end{align}
where the operator ${\cal F}_l(\cdot)$ flipped $\sigma_l^t$ to $-\sigma_l^t$.
The difference of the Hamiltonian 
$H(\Vec{\sigma}^t) - H({\cal F}_l(\Vec{\sigma}^t))$ 
can be denoted as
\begin{align}
 & H(\Vec{\sigma}^t) - H({\cal F}_l \Vec{\sigma}^t) = 2 \sigma_l F_l(\Vec{\sigma}^t)\\
 & F_l(\Vec{\sigma}) = \frac{1}{(r-1)!} \sum_{i_2, \cdots i_r} 
 J_{l i_2 \cdots i_r} \sigma_{i_2}^t \cdots \sigma_{i_r}^t + h \tau_l
\end{align}
This is equivalent to determining the $l$th site probability by
\begin{equation}
 P(\sigma_l^{t+1} = \pm 1) = \frac{1}{2} \pm 
 \frac{\tanh (\beta F_l( \Vec{\sigma}^t ))}{2}.
\end{equation}
In this simulation, 
the initial state $\Vec{\sigma}^1$ was set to the true message $\Vec{\xi}$, 
so the simulation result indicated the stability of the true message
in the model.

In the analog model simulation
we used the synchronous update rule describe in Eq..(\ref{eq:dynamics}).
Thus all units were updated simultaneously.
In the analog simulation, we also set the true message $\Vec{\xi}$ 
as the initial state $\Vec{s}^1$.

Figs.\ref{fig:SimResult}(a) and (c) show the binary model 
simulation results along with the theoretical analysis results
at $j_0/J=0.8$ and $j_0/J=1.0$, respectively.
(The horizontal axis shows  $T/J$, and the vertical axis shows the 
absolute value of the overlap $|M_o|$.)
The computer simulation results are represented by error bars showing the quartile
deviation, and the dashed line shows the theoretical analysis result.
The phase transition occurred at about $T/J = 1.0$ and $T/J=1.5$ 
in figs.\ref{fig:SimResult}(a) and (c), respectively,
and the results of the computer simulation agree with the theoretical
analysis results.

Figs.\ref{fig:SimResult}(b) and (d) shows the analog model simulation
results along with   the
theoretical analysis results at $j_0/J=0.8$ and $j_0/J=1.0$, respectively.
The computer simulation results are again represented by error-bars, and 
agree with the replica analysis results.
In figs.\ref{fig:SimResult}(b) and (d), 
the theoretical results show that the phase transition occurred at about
$T/J=1.5$ and $T/J = 1.9$, respectively.
The computer simulation agreed with the analysis at such high temperatures.
In the analog model, 
the critical temperatures, which cause the phase transition, 
became higher than those of the binary model.
This indicates that
the deterministic analog decoding model is more robust than 
the binary decoding model in terms of decoding ability
when the receiver overestimates the decoding temperature.

However, 
at low temperatures in the analog model % shown in fig.\ref{fig:SimResult}(b), 
the simulation results did not agree with theoretical result 
(Fig.\ref{fig:SimResult}(b)).
%In such a low temperature region ($T/J \leq 0.55$), 
%the equilibrium state is considered the `spin glass' state ($m=0$ and $q>0$). 
The simulation initial state is $\Vec{s}^1 = \Vec{\xi}$,
and there exist meta-stable states around the $\Vec{\xi}$,
so 
the dynamics of $\Vec{s}^t$ is captured by this state and 
the absolute overlap $|M_o|$ stays close to $1.0$.

An important feature of 
the analog model simulation results shown in
figs.\ref{fig:SimResult}(b) and (d),
is  that
the absolute value of the overlap $|M_o|$ was exactly $0$ at high temperature.
In the analog model, 
the all units value $s_i$ exactly converged to $0$ for any messages $\{\Vec{\xi}\}$,
so  
we can guess whether the retrieval is failed or not
at high temperature exceeding the retrievable limit.
In practical,
the ability to determine whether
decoding will be finished in failure or success
is a desirable feature.

\subsection{Convergence Speed}
\begin{figure}
 \begin{center}
  \resizebox{0.45\textwidth}{!}
  {\includegraphics{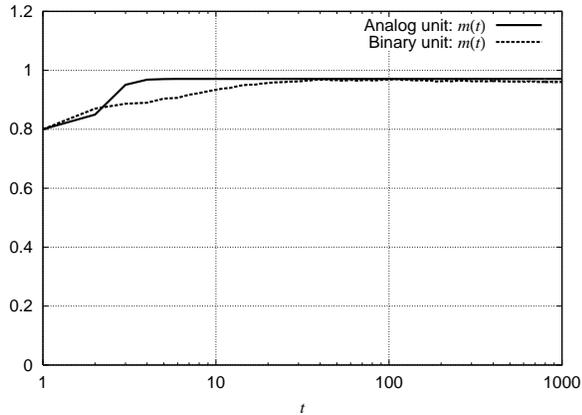}}
 \end{center}
 \caption{Convergence of the macroscopic quantity $m(t)$: the solid
 line represents the analog model and the dashed line represents the binary
 model. $t_{\text{min}}=1$
 }
 \label{fig:sim_macroconv}
\end{figure}

The main advantage of the NMF approximation is the low calculation cost of convergence.
Here,
we discuss the convergence time that should be set in the computer simulation for each method.
To calculate the thermal average in the computer simulation, 
we implemented
\begin{align}
 \overline{\langle \sigma^t_i \rangle} &= \frac{1}{T}
 \sum_{\tau = t_{\text{min}}}^{t} \sigma_i^{\tau}, \\
 T &= t - t_{\text{min}}+1,
\end{align}
as the thermal average, 
where the superscript $t$ means discrete time and
$t_{\text{min}}$ means the beginning of thermal average calculation.
%
%
%We observed convergence time of the overlap in each simulation.
First, we observed the macroscopic quantities $m^t$;
\begin{equation}
 m^t = \left\{ 
	\begin{array}{ll}
	\frac{1}{N} \sum \overline{\langle \sigma^t_i \rangle} & \:\:\: {\text{(binary model)}} \\
	\frac{1}{N} \sum s^t_i       & \:\:\: {\text{(analog model: NMF approximation) }}
	\end{array}
       \right.
\end{equation}
In this simulation, 
we set the initial value $\Vec{s}^1$ and $\Vec{\sigma}^1$ 
to  satisfy the overlap $M_o = 0.8$, and $t_{\text{min}} = 1$.
Fig.\ref{fig:sim_macroconv} shows the dynamics of $m^t$.
The solid line represents $m^t$ of the binary stochastic model and
the dashed line represents that of the analog deterministic model.
The convergence times for the two lines seem to be of the same order.
%In each line, it looks the converge time is same in same order.

\begin{figure*}[t]
 \begin{tabular}{cc}
 \resizebox{0.45\textwidth}{!}
 {\includegraphics{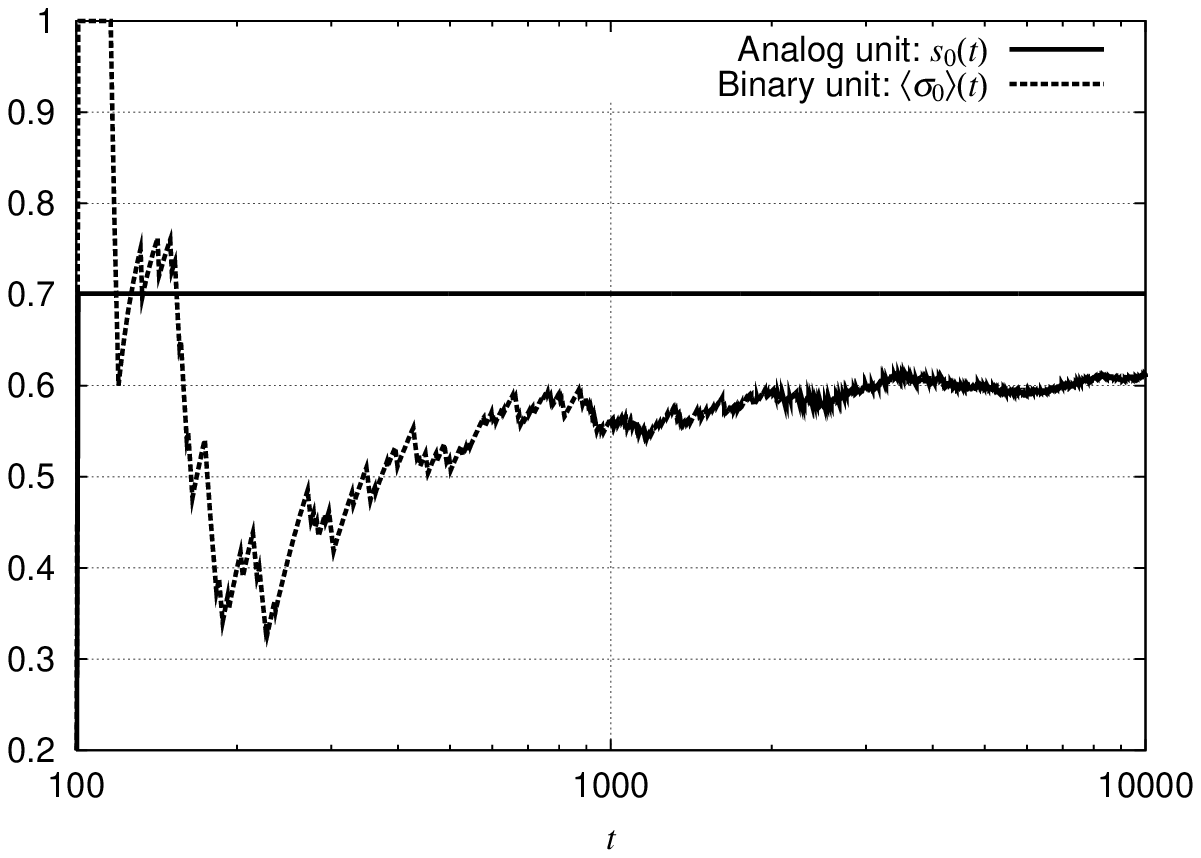}}
  &
 \resizebox{0.45\textwidth}{!}
 {\includegraphics{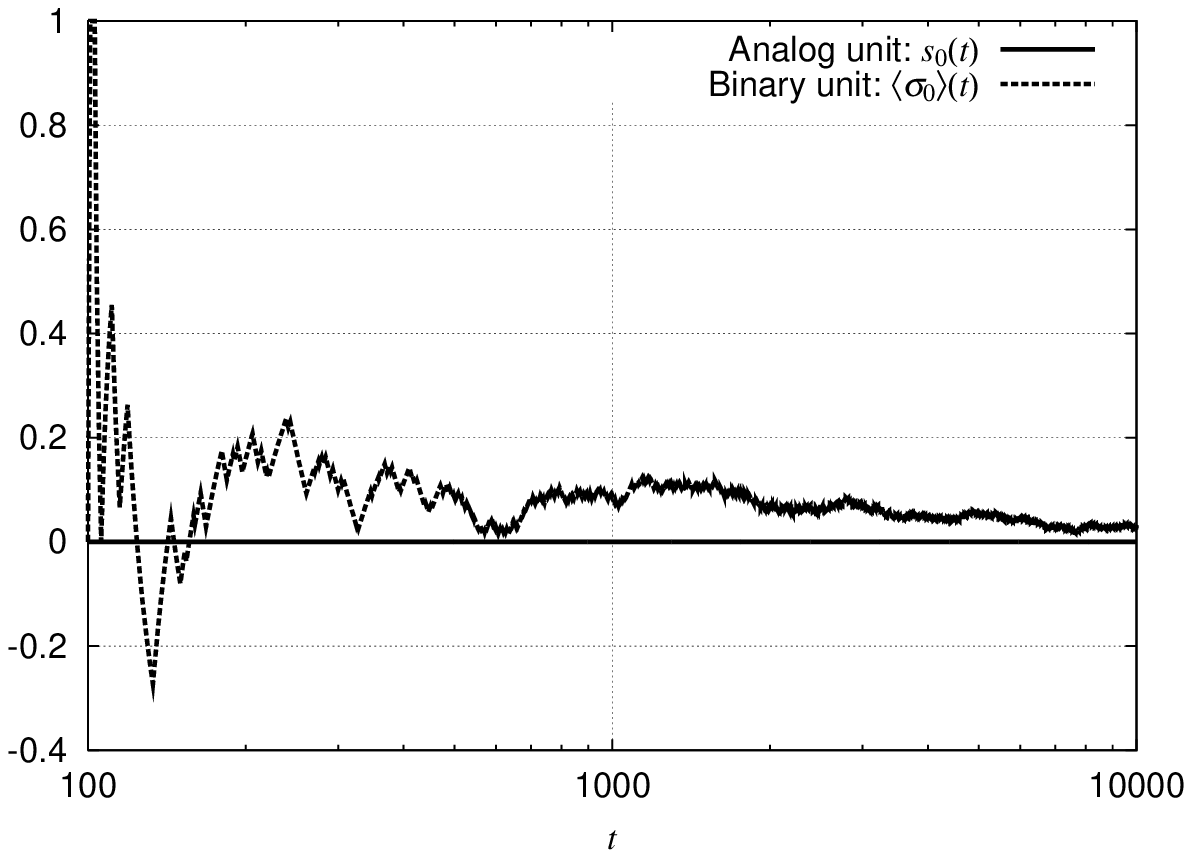}}
  \\
  (a) & (b)
 \end{tabular}
 \caption{
 Typical dynamics of $\overline{\langle \sigma_1^t \rangle}$ and $s_1^t$.
 The solid lines show the dynamics of $s^t_1$ which converged within
 $t_{\text{min}} = 100$ iterations. 
 The dashed lines show the dynamics of  
 $\overline{\langle \sigma_1^t \rangle}$.
 In (a), the S/N ratio and temperature parameters were set to 
 $j_0/J = 1.0$ and $T/J = 1.0$, respectively.
 The convergence of $\overline{\langle \sigma_1^t \rangle}$ was slower
 than that of $s^t_1$.
 In (b), the temperature was set to 
 $T/J = 1.80$; i.e., outside of the retrieval region.
 In this case the convergence of $\overline{\langle \sigma_1^t \rangle}$ was also slow.
 }
 \label{fig:sim_conv}
\end{figure*}
In the MPM inference, however, we should determine each unit's value;
that is, the microscopic quantity.
Thus, we investigated the behavior of the thermal average of unit
$\overline{\langle\sigma_1^t \rangle}$ 
for the binary model and $s_1^t$ for the analog model. 
In the analog model, all units were updated simultaneously because 
we adopted synchronous updating. 
In contrast, we adopt asynchronous updating in the binary model.
To compare the convergence time between analog unit and binary unit,
we regarded $N$ updates as one Monte Carlo step ($1$ MCS) for binary model,
where $N(=100)$ means the number of units in the simulation.
Thus, one MCS update corresponds to one synchronous update in the
analog model.
Fig. \ref{fig:sim_conv}(a) shows a typical result regarding the convergence speed for
the S/N ratio $j_0/J = 1.0$ and temperature $T/J=1.0$;
we set the beginning of the thermal average calculation 
as $t_{\text{min}}=100$ MCS.
The horizontal axis shows the calculation time measured by MCS, and 
the vertical axis shows the value of $\sigma_1^t$ and $s_1^t$.
The solid line shows typical dynamics of $\sigma_0^t$, and the dashed line
shows typical dynamics of $s_0^t$.
In macroscopic points of view, each model achieved the same overlap $|M_o|$, 
but in microscopic perspective, the deterministic analog model converged
about over 1000 times as quickly as the stochastic binary model.

Fig. \ref{fig:sim_conv}(b) shows the dynamics at a higher temperature $T/J = 2.2$.
$\overline {\langle\sigma^t_1 \rangle}$ did not converge to 0; however
$s_1^t$  seemed to be converging to exactly $0$ in the early time steps.
Thus, when using the analog model, we can easily predict within the
early time steps whether 
the decoding will be finished in failure or not.

\section{Conclusion}
In this research, we have investigated an error-correcting code that
uses the MPM inference.
Since the MPM inference requires many trials to calculate
the thermal average for each unit,
we tried to replace this operation with a form of deterministic 
analog dynamics called
naive mean field approximation.
We analyzed the decoding ability with the deterministic analog model 
through the replica method, 
and quantitatively compared it with that of the stochastic binary model 
suggested by Nishimori \& Wong\cite{Nishimori00} .
We found that the decoding ability of the deterministic analog model is superior to 
that of the stochastic binary mode l
at higher temperature area.
To confirm this result, we carried out a  computer simulation for each
model and obtained the results that agreed with our analysis results.

\section*{Acknowledgement}
This work was partially supported by research grant No.15700192 of the
Grant-in-Aid for Young Scientists (B) in Japan.

%\bibliographystyle{unsrt}
%\bibliography{shouno}

\end{document}